\documentclass{eas}
\usepackage{graphicx}
\usepackage{MR_eas}
\usepackage{natbib}
\begin{document}

%%-----------------------------
%%      the top matter
%%-----------------------------
\title{Modeling rotating stars in two dimensions} 
\author{Michel Rieutord}
\address{Universit\'e de Toulouse; UPS-OMP; IRAP; Toulouse, France
and CNRS; IRAP; 14, avenue Edouard Belin, F-31400 Toulouse, France}
\begin{abstract}
In this lecture I present the way stars can be modeled in two dimensions
and especially the fluid flows that are driven by rotation. I discuss
some of the various ways of taking into account turbulence and conclude this
contribution by a short presentation of some of the first results
obtained with the ESTER code on the modeling of interferometrically
observed fast rotating early-type stars.
\end{abstract}
\maketitle
%%-----------------------------
%%      your text
%%-----------------------------
\section{Introduction}

Rotation remains, with magnetic fields, an ill-known quantity in the
interiors of stars. It is however associated with fascinating objects
like massive stars or the first stars of the Universe. These latter objects
are indeed often called the factories of metals and because of their
compactness (due to low opacity) are usually thought to have been
fast rotators. However, understanding the mechanisms that lead to the
enrichments of the interstellar medium with the variety of elements,
requires the understanding of the mechanisms which mix the stars or
which simply bring the elements from the regions of their nucleosynthesis to
the surface where they can be expelled by winds.

The determination of the mean flows that pervade rotating stars is
therefore an unavoidable step towards this understanding. This is why
much work has been devoted to insert the effects of flows into one
dimensional models \cite[e.g.][]{MM00,maeder09}. Even if this modeling
succeeded in explaining a variety of effects of rotation (e.g. the ratio
of blue to red supergiants in galaxies or the abundance of lithium), 1D
models cannot include many specificities of rotating fluid flows.
Indeed, a fluid flow is intrinsically a multidimensional phenomenon. Two
dimensions of space are therefore the minimum number of dimensions
to compute a rather general flow. For instance,
the geostrophic flow of rotating fluids, that comes from
by the domination of the Coriolis force, verifies
the Taylor-Proudman theorem stating that $\dz{\rho\vv}=\vzero$. Such a
condition is not compatible with the existence of a stellar core. In
fact the very problem is that rotation imposes a cylindrical symmetry
while gravity imposes spherical symmetry. The combination of both yields
a two dimensional problem.

In the following I therefore propose to focus our attention on the mean
flows that pervade a rotating star. We will thus have the opportunity to
go through the processes (baroclinicity and Reynolds stresses) that
drive a secular mixing of the stars. Then, we will shortly present the state
of the ESTER project and some selected first results on the modeling of
early-type fast rotating stars.

\section{The problem's formulation}

In order to make a first step into the 2D-modeling of rotating stars,
we shall consider an isolated, non-magnetic, not-mass-losing, early-type
star. Moreover, we shall forget about any time evolution: the star is
powered by nuclear reactions but these do not influence the chemical
composition. However, there are some regions with turbulent flows. There we
assume that the turbulence is in a statistically steady state. Such an
ideal star does not exist but some stars like Vega may be close to it.

With all the foregoing precautions, we may formulate the problem of a
2D-model of a rotating star in a consistent way.

The partial differential equations that govern the steady-state of our
ideal rotating star read:

\greq
     \Delta\phi = 4\pi G\moym{\rho} \\
     \moym{\rho T \vv\cdot\na S} = -\Div\vF + \eps_*\\
     \moym{\rho \vv\cdot\na\vv} = -\na \moym{P} -\moym{\rho}\na\phi+\vF_v\\
     \Div\moym{\rho\vv} = 0.
\egreqn{basiceq}
These are Poisson equation for the gravitational potential $\phi$,
the energy equation involving the temperature $T$, the density $\rho$,
the entropy $S$, the diffusive heat flux $\vF$ and the nuclear heat
sources $\eps_*$, the momentum equation with the viscous force
$\vF_v$ and the equation of mass conservation. The brackets $\moym{...}$
indicate time averages. They are necessary to remove turbulence
fluctuations. Actually, all quantities are time averaged.

Equations \eq{basiceq} should be completed by the prescriptions of the
microphysics, namely the equation of state, the opacities, the nuclear
network etc. They should also be supplemented with boundary conditions.
All these are discussed in \cite{ELR13}. Here, we shall only focus on
the problems raised by the velocity and temperature fields boundary
conditions. Before that we need to concentrate on the mean flows.

\section{The mean flows in rotating stars}

The mean flows that pervade a rotating star and enforce some mixing
have various origin. The most common sources are the following:

\begin{itemize}
\item baroclinicity
\item Reynolds stresses
\item gravitational contraction, mass loss or mass accretion
\end{itemize}
Let us review these phenomena in more details.

\subsection{Baroclinicity}\label{barocsect}

In fluid mechanics baroclinicity refers to the inclination of isobars
and isotherms (or isopycnic, equipotential surfaces...). This is indeed a
natural feature of rotating stars that was discovered long ago by
\cite{vonzeipel}. Since much confusion has emerged after this work on
the origin of the meridional circulation, I'd like to say a few words
about stellar baroclinic flows in order to
clarify the phenomenon and the way it should be approached.

Baroclinicity emerges in fluids because pressure and temperature obey
two different and independent equations: there are some coupling
through buoyancy and heat advection but these are weak. Hence, if we think
to these two fields (pressure and temperature), they live independently
and therefore establish there own system of isosurfaces.  Usually, they do
not coincide and baroclinicity arises. It arises because density depends
on temperature or on both temperature and pressure. Hence, isodensity
surfaces are distinct from isobars or, in other words, the two vectors
$\na\rho$ and $\na P$ are not parallel.  When taking the curl of the
momentum equation (after division by $\rho$), one gets

\[ M(\vv) = \frac{\na P\times\na\rho}{\rho^2}\]
where the right hand side is usually called the baroclinic torque and $M$
is a differential operator. The consequence of this torque is that a flow
arises. In stars this flow is basically a differential rotation, namely a
purely azimuthal velocity field. However, as viscosity is always present,
a differential rotation, like any shear flow, transport momentum in the
direction of the velocity gradients. In a steady state the momentum flux induced
by viscosity must be compensated by some meridional flow. Such a balance
is illustrated by the $\varphi$ component of the momentum equation,
which may be written

\[ v_s\partial_s(sv_\varphi) + v_z\partial_z(sv_\varphi) = \nu
s\Delta'v_\varphi \]
or, in a more condensed form,

\beq \vv\cdot\na(s^2\Omega) = \nu \na\cdot(s^2\na\Omega)\eeqn{advdiffl}
where we used the cylindrical coordinates $(s,\varphi,z)$ and where
$\Delta'$ represents some Laplacian-like operator.

As we may note on \eq{advdiffl}, advection of angular momentum
$\vv\cdot\na(s^2\Omega)$ just compensates the viscous force due to angular velocity
gradients $\nu \na\cdot(s^2\na\Omega)$. In
axisymmetric situations like that of our rotating stars, advection is that of
meridional circulation. We therefore see that in a steady state
meridional circulation is controlled only by viscosity.

Confusion arose in the past
because it was assumed that rotation generates a thermal
disequilibrium. Indeed, in a barotropic fluid

\[ \Div(\khi\na T) \neq 0\]
known as von Zeipel's paradox. The paradox was solved  by saying that 
the meridional circulation velocity $\vv$ was such that

\[ \rho T\vv\cdot\na S = \Div(\khi\na T) \andet \Div(\rho\vv) = 0\]
namely, heat advection by meridional circulation compensates the thermal
imbalance and verifies mass conservation. {\em Such a reasoning is
incorrect} because the driving of a flow is due to forces or torques,
not to mass and energy conservation which cannot ensure angular momentum
conservation. As shown by \cite{Buss81}, a thermal imbalance leads to
a time-evolution of the temperature field itself, which is more rapid
than a mechanical rearrangement of the fluid.  The foregoing discussion
is detailed in \cite{R06b}.

\subsection{Reynolds stresses}

\subsubsection{The scale of turbulence}

Baroclinic flows therefore slowly advect heat, chemicals and angular
momentum.  However, the differential rotation of baroclinic origin
is a mere shear flow that may develop instabilities if the Reynolds
number is large enough or if the Richardson number is low enough. These
instabilities usually lead to turbulence. Unlike thermal convection where
the linear instability drives motion at scales on the fraction of the
radius, shear instabilities in a stably stratified fluid inject energy
at small-scale. To see that, we recall that the Richardson criterion
taking into account the high thermal diffusivity of the fluid imposes
\cite[][]{zahn92}

\[ \frac{N^2}{S_h^2}\frac{v\ell}{\kappa} \infapp 1/4\]
where $N$ is the \BVF, $S_h$ the local shear, $v$ and $\ell$ the velocities and
length scale of the eddies and $\kappa$ the heat diffusivity. This criterion
originally proposed by \cite{townsend58}, says that shear instability grows
up to a scale (of the eddies) where heat diffusion is unable to smooth
out the stable buoyancy of the
background. Since the largest eddies have a turn-over time scale $1/S_h$, we may
observe that all the eddies forced by the background shear are such that

\[ \frac{N^2}{S_h^2}\frac{\ell^2}{\tau}\infapp \kappa/4, \quad {\rm with}\quad
\tau\infapp S_h^{-1}\]
or

\[ \frac{\ell}{R} \infapp \lp\frac{\kappa S_h}{4R^2N^2}\rp^{\!1/2}\]
where we introduced $R$ the radius of the star. Let us put numbers in this
expression. From \cite{ELR13}, we evaluate the associated turbulent
kinematic viscosity

\[ \nu_t = \frac{S_h^2\kappa}{12N^2} \sim 10^6 \; {\rm cm}^2/s\]
The global shear of differential rotation is of a fraction of the 
rotation rate; let us write $S_h=f\Omega$. Hence we find

\[ \frac{\ell}{R} \infapp
\lp\frac{\nu_t}{4fR^2\Omega}\rp^{\!1/2}=\frac{\sqrt{E}}{\sqrt{2f}}\]
where $E$ is the Ekman number based on the shear induced turbulence. Since
$E< 10^{-8}$ and $f\sim 0.05$ \cite[see][]{ELR13}, we find that

\[ \ell/R \infapp 3\times 10^{-4}\]
Hence, turbulence in vertically stratified region remains on relatively
small-scales in the vertical direction. In the horizontal directions of
course the scales can be large (but limited by the stability of the
eddies).

As far as convection zones are concerned, nothing prevents the
instability from driving all the scales where the Schwarzschild criterion
predicts convection.  Hence these are fully mixed.

\subsubsection{Reynolds stresses}

The convection zones are fully mixed regions, but because of the
strong turbulence they harbour, Reynolds stresses drive mean flows like
the differential rotation of the Sun. These stresses and the associated
flows also force some flows in the neighbouring radiative zones. The
picture that emerges from these remarks is that turbulence is everywhere
in stars, with variable intensity, properties, and a modeling of its
effects is required. Here we first concentrate on its momentum transport
that is on Reynolds stresses. If we use Reynolds decomposition, namely

\[ \vv = \moym{\vv} + \vv'\]
where $\vv'$ are the velocity fluctuations with respect to the average, the Reynolds
stress tensor is 

\[ {\bf R} = \moym{\rho\vv'\otimes\vv'}\]
Usually, correlations with density are not important (a fortiori in an
incompressible fluid!) and the components of the Reynolds stress tensor
are often written

\[ R_{ij} = \moym{v'_iv'_j}\]
In a steady state, and neglecting correlations with density, mean flows therefore
verify:

\[ \rho\moym{v_j}\partial_j\moym{v_i} + \partial_j(\rho\moym{v'_iv'_j}) =
-\partial_i\moym{P} + \rho g_i + F_i\]
where $\vF$ is the viscous force.

The main question with the previous equation is the expression of the Reynolds
stress tensor as a function of the mean-field, the so-called closure problem of
mean field equations. 

A basic way of solving this problem is to assume that turbulent stresses
are like fluid stresses and that they can be represented by a turbulent
viscosity. We thus may write

\[ \rho\moym{v_j}\partial_j\moym{v_i} =
-\partial_i\moym{P} + \rho g_i + F^{\rm T}_i\]
with 

\beqan
\vF^T &=&
\displaystyle\mu_t\lc\Delta\vv+\frac{1}{3}\na\left(\na\cdot\vv\right)
+2\left(\na\ln\mu_t\cdot\na\right)\vv\right.\nonumber \\
&&\displaystyle
\quad\left.+\na\ln\mu_t\times(\na\times\vv)
-\frac{2}{3}\left(\na\cdot\vv\right)\na\ln\mu_t\rc \;.
\eeqan{fvisc}
where $\mu_t$ is the turbulent dynamical viscosity of the gas. This
expression enables local variations of the viscosity although it does
not say anything on its determination. But there is
a more fundamental difficulty: there is no reason why the functional
form of the Reynolds stress should be

\beq \moym{\rho v'_iv'_j} =
-\mu_t(\partial_i\moym{v_j}+\partial_j\moym{v_i}
-\frac{2}{3}\partial_k\moym{v_k}\delta_{ij})\eeqn{visct}
as expected for a newtonian fluid. The analogy of turbulence with such a
fluid is very limited because this expression is derived from the
assumption that fluid flows are
slight perturbations of the thermodynamic equilibrium. It is unclear what is
the statistical equilibrium of turbulence and how actual flows may slightly
deviates from it.  Moreover, turbulence particles are vortices that
support long range interactions, not collisions...

Now, if we still admit \eq{visct}, the expression of $\mu_t$ is still
a problem. In the nineteentwenties, Prandtl introduced the idea of the
mixing length suggesting that in turbulent shear flows

\[ \mu_t = \rho\Lambda^2\left|\dz{v_x}\right|\]
where $\Lambda$ is the mixing length and $z$ the direction of the
shear. This assumption leads to interesting results on wall-turbulence,
but misses the point on the axis of a turbulent jet (where it predicts
zero-viscosity!).

Another recipe is the Smagorinski's prescription saying that
$\mu_t=\rho\Delta^2|\moym{c_{ij}}|$ where $\moym{c_{ij}}$ is the mean
shear. Such a prescription is more general and still often used in
Large-Eddy Simulations (LES) as a basic subgrid model. However, as the
mixing-length model, this is a local prescription that does not take
care of non-local effects.

To avoid this pitfall, more sophisticated models have been developed for
engineering problems. One of the most popular is the $K-\eps$ model of
\cite{LS72}: this models adds two new equations for the evolution of $K$
the turbulent kinetic energy and $\eps$ the energy dissipation. The
advection diffusion equation that controls the evolution of these
quantities are derived from the evolution of second-order correlations
with a simple model of the third-order correlation \cite[][]{R97}. The
point is that when $K$ and $\eps$ are known, the turbulent pressure
and the turbulent viscosity are also known. These models may be useful
to determine statistically steady inhomogeneous flows that we find in
stars. However, because of the huge increase of the computing power,
these models have been forsaken to the profit of LES, which are more
flexible for industrial flows. In astrophysics, this raises the problem
of the comparison with observations, which cannot be as detailed as
laboratory experiments.

Another line of research that should be mentioned, is the one followed
by R\"udiger and collaborators \cite[see][]{Rudig89,KPR94}. This is the
mean-field approach of turbulence where one seeks for an expression of
$\moym{v'_iv'_j}$ and other second order correlations, as functions
of large-scale quantities like local rotation rate. In this approach
the Reynolds tensor is not reduced to that of the functional form of a
newtonian fluid. New effects like the $\Lambda$-effect or anisotropic
turbulent viscosities, appear. These approaches of course raise new
kinds of problems \cite[see][]{snellman_etal12}.

\subsection{Gravitational contraction, mass accretion and mass-loss}

The last source of large-scale motions in a rotating star is due to
expansion or contraction of the star at various phase of its
evolution. In non-rotating stars these phenomena just lead to
radial flows, but in a rotating star, angular momentum conservation
leads to a differential rotation and an associated meridional
circulation. Not much is known on these flows that are currently
under investigations (Hypolite \& Rieutord in preparation). Again, the
associated shear will lead to some small-scale turbulence in radiative
regions.

\subsection{Conclusion}

To conclude the foregoing discussion, let us underline that we
considered here only non-magnetic processes. Magnetic fields will of
course complicate the picture. An open question is how all these
sources compete together and of course how we should model the associated turbulence.
%transport is still an open question rooted to the general problem of
%turbulence modeling.  Thus any stellar structure code should leave open
%the possibility of updating this side of the model as our understanding
%of turbulence progresses.

\section{Heat transport}

\subsection{convection zones}

The foregoing discussion focused on turbulent viscosity and more generally
on momentum transport. However, heat transport was really the first
obstacle to circumvent when first stellar models were constructed. The
well-known mixing-length theory, based on Prandtl's idea, offers a
plausible answer to this problem. In this approach, all the difficulties
of turbulent heat transport are condensed in a dimensionless parameter
of order unity $\alpha_{\rm MLT}$ that has been calibrated on the
Sun. Unfortunately, such a parameter is not universal and varies from
star to star, impeding any precise prediction. A clear example is given
by the two stars of $\alpha$ Cen \cite[][]{eggenberger_etal04}.

In two dimensions, the difficulty is squared. First because the 1D-MLT
has no straightforward generalisation to 2D. Turbulent transport is now
also in latitude via turbulent diffusion and mean flows.

A first idea is use the ``down-gradient" prescription for the heat flux
(this is a prescription that is largely used in the $K-\eps$ model). Here,
since thermal convection disappears when the entropy gradient vanishes
or becomes positive (in the direction opposite to effective gravity),
a natural way to model the convective heat flux is to assume that it is
proportional to the entropy gradient, namely

\[ \vF_{\rm conv} = \moym{\rho c_pT\vv} = -\khi_{\rm turb}T\na S/\calR\]
where $\calR$ is the ideal gas constant. 

When reduced to 1D this prescription can be related to the
classical MLT. It may therefore be viewed as a generalization of this
theory. However, there are other possibilities like the introduction
of an eddy conductivity $\khi_t$ such that $\vF=-\khi_t\na T$
\cite[][]{Rudig89}.

\subsection{The surface heat transfer}

One of the new problem raised by 2D-models concerns the implementation of the
boundary conditions that should be met by the temperature field. A simple
prescription that replaces a detailed atmospheric model is to impose that the
star radiates locally as a black body, namely that

\beq -\khi_r\vn\cdot\na T = \sigma T^4 \eeqn{bctemp}
where $\khi_r$ is the radiative conductivity, $\sigma$ Stefan-Boltzmann
constant and $\vn$ the outer normal of the surface.  In non-rotating
models this condition is usually applied where the optical depth
is $\tau_s=2/3$, but $\tau_s=1$ is also used \cite[][]{Morel97}. In
2D-models, the situation is more complicated: one has to define the
surface where boundary conditions (for velocity, gravitational potential
and temperature) are imposed. Since the opacity is a rapidly varying
quantity at the surface of a star an iso-optical-depth surface is not
convenient numerically. We therefore chose to impose boundary conditions
on an isobar and select the isobar that ``emerges" at the pole, i.e. whose
pressure is

\beq P_s=\tau_s\frac{g_{\rm pole}}{\kappa_{\rm pole}},\eeq
On this isobar $T=T_{\rm eff}$ at the pole only. As one moves towards
the equator this isobar sinks into the optically thick matter because at
$\tau_s=2/3$ pressure is stronger at the pole than at the equator (recall
that gravity is larger at the pole). The trick here is to determine the
temperature on this isobar as a function of the latitude. For this we
assume that the matter lying above this isobar behaves like a polytrope
of index $n$. With this hypothesis, it can be shown \cite[cf][]{ELR13}
that the temperature on this isobar follows the law

\beq T_b(\theta) = \lp\frac{g_{\rm pole}}{g_{\rm
eff(\theta)}}\frac{\kappa(\theta)}{\kappa_{\rm
pole}}\rp^{1/(n+1)}\lp\frac{-\khi_r\vn\cdot\na T}{\sigma}\rp^{1/4}
\eeqn{tb}
which fully determines the temperature field. In more sophisticated
models, $T_b(\theta)$ may be determined by the modeling of the stellar
atmosphere.

\section{Computing the baroclinic flows in 2D-models}

We focus on early-type stars so as to avoid the modeling of
outer convective layers. Thermal convection is only modeled in the
core of the star.  One-dimensional models tell us that it is very
efficient and therefore the assumption of an isentropic core is quite
good \cite[][]{maeder09}. ESTER models therefore assume isentropic cores
and no further modeling of core turbulence is included yet.

\subsection{Flows in the radiative envelope}

Above the core, the radiative envelope is not in hydrostatic equilibrium
as recalled above. Baroclinic flows pervade it and are responsible of
the mixing.  They represent the first difficulty for the computation
of self-consistent models.  When determined, we have a prediction of the
differential rotation and the meridional circulation.

As we noted in sect.~\ref{barocsect} the determination of these
flows rests on the fluid's viscosity, which is crucial for stationary
solutions. The importance of viscosity is shown by the value of the
Ekman number, namely

\[ E = \frac{\nu}{2\Omega R^2}\]
As shown in \cite{ELR13}, its value is always less than 10$^{-8}$
even if shear turbulence is accounted for.

Viscous effects are therefore expected to be small. But they are crucial
as they lift the degeneracy of the differential rotation as we shall
show now.

Let us take the curl of the momentum equation and let us project it in
the azimuthal direction on $\ephi$. If viscous or Reynolds stresses are
neglected then one finds:

\beq s\dz{\Omega^2} = \frac{\na
P\times\na\rho}{\rho^2}\cdot\ephi\eeqn{baroc}
The solution of \eq{baroc} is invariant with respect to the
addition of an arbitrary function of $s$; if $\Omega(s,z)$ is a solution

\[ \Omega'^2(s,z) = \Omega^2(s,z)+F(s)\]
is also a solution. As shown in \cite{R06}, viscosity lifts this
kind of degeneracy. This may be understood in the following way. The
differential rotation depends on $F(s)$. The balance of angular momentum
advection and diffusion is given by \eq{advdiffl} which is completed by
mass conservation. Once $\Omega$ is known, these two equations give the
meridional circulation, but without matching the boundary conditions. In
general, the derived meridian flow does not verify 

\[ \vv\cdot\vn =0\]
on the stellar surface. This condition is enforced by an Ekman boundary
layer where the mass flux into the surface is identified with the
mass pumping of the boundary layer. Indeed, the Ekman boundary layer
completes an inviscid solution like $s\Omega(s,z)+G(s)$ in such a way
that stress-free conditions are met, but the boundary layer correction
$\tu_\theta,\tu_\varphi$ usually do not verify mass conservation. This
latter constraint is corrected by the velocity component orthogonal to the
layer, which is called the pumping. The identification of the pumping of
the layer with the mass flux generated by the circulation at the stellar
surface gives the differential equation for the unknown geostrophic
flow that appears in inviscid solutions ($F(s)$ or $G(s)$). This
differential equation is derived in \cite{R06}, in the simplified case
of an incompressible ``star" not far from rigid rotation.

In modeling rotating stars in 2D we are not interested in the boundary
layer because they are very thin and would require a high spatial
resolution near the surface. This is why \cite{ELR13} derived a
special boundary condition that completes \eq{baroc} and insure
stress-free conditions without an explicit computation of the boundary
layer flow. This boundary condition reads

\begin{equation}
\label{eq:bl}
E_s
s^2\vec{\hat\xi}\cdot\na\Omega+\psi\vec{\hat\tau}\cdot\na(s^2\Omega)=0
\qquad\mbox{at the surface}\;.
\end{equation}
where $E_s$ is the surface Ekman number, $\vec{\hat\xi}$ and
$\vec{\hat\tau}$ unit vectors normal and tangential to the surface,
while $\psi$ is the stream function of the meridional circulation. It turns
out that this circulation scales as the interior Ekman number, therefore
this small parameter drops out of the momentum equation and only order
one quantities are computed.

\subsection{The Core-Envelope Interface}

The foregoing simplification unfortunately breaks down at the
core-envelope interface (CEI). Indeed, as shown by \cite{ELR13} the chemical
evolution of the convective core, that leads to a density discontinuity
at the CEI, also leads to an angular velocity discontinuity on this
interface. In addition we may also consider that turbulent viscosity
steeply increases when one goes from the envelope to the core. These
discontinuities are triggering the so-called Stewartson layer that
develops along the tangential cylinder circumventing the core. This
layer may play a crucial role in the chemical enrichment of the envelope
by products of the nuclear reactions. Its computation is however demanding
as it scales like E$^{1/4}$ but fortunately less than the Ekman layer
whose thickness scales like E$^{1/2}$. More studies are necessary to
better determine the properties of this interfacial region, which is also
suspected of harbouring some convective overshooting.

\begin{figure}[t]
\includegraphics[width=\textwidth]{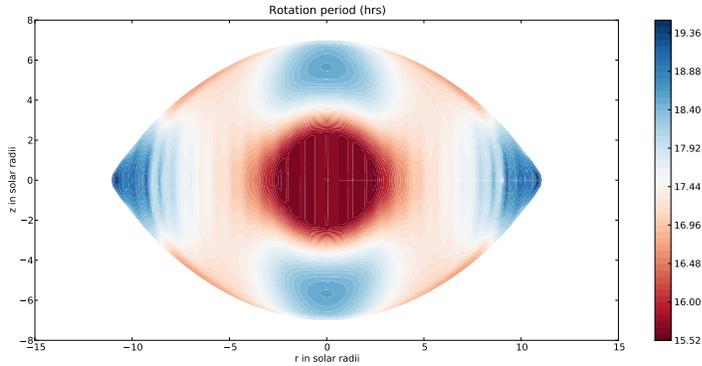}
\caption{Meridional view of the differential rotation of 30~\msun\ ZAMS star
rotating at 98\% of its critical angular velocity with X=0.7 and
Z=0.02.}
\label{Massive}
\end{figure}

\section{Some results of the ESTER project}

A detailed account of the present achievements of the ESTER project may be
found in the two papers by \cite{ELR13} and \cite{REL12}. Here, we wish
to give a brief summary of the performance and first results of the
ESTER code\footnote{The ESTER code is a public domain code that is
freely available at \tt http://code.google.com/p/ester-project/}.

\begin{figure}[t]
\includegraphics[width=\textwidth]{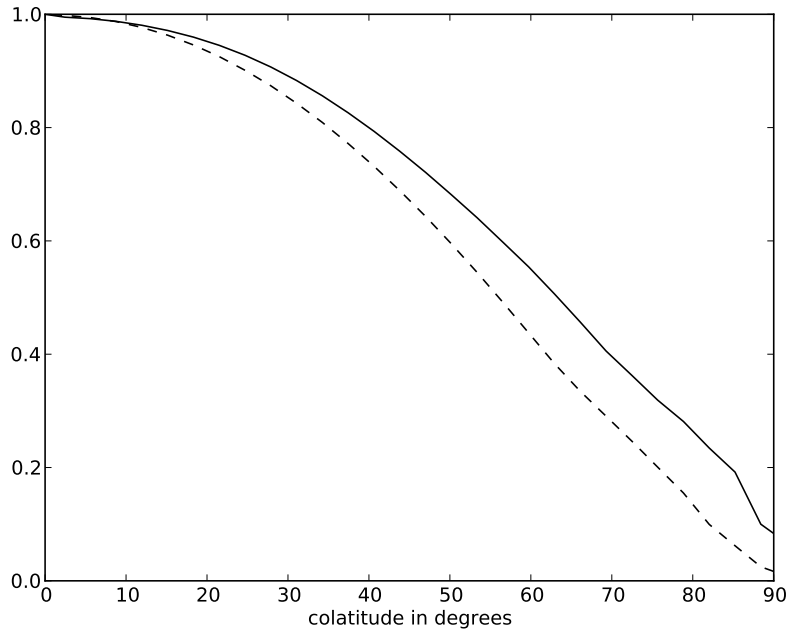}
\caption{Variations of the flux with colatitude for a ZAMS star of 30~\msun\
rotating at 98\% of its critical angular velocity. Solid line: the ratio
of the flux to the polar flux as a function of the colatitude. Dashed
line: the prediction of the von Zeipel law.}
\label{flux}
\end{figure}

\subsection{Presentation}

The ESTER code is solving the equation of stellar structure in two
dimensions \eq{basiceq}, assuming an isentropic convective core and
a radiative envelope, thus restricted to the modeling of an early-type
star. We use OPAL tables for the equation of state and
opacities. Convection in layers is not computed yet : the temperature
gradient is assumed to be close to the radiative one, which is fine for
stars with mass larger than 1.8~\msun.

The discretization of the PDE is based on a spectral decomposition:
Chebyshev polynomials radially and spherical harmonics horizontally. The
distorted shape of the star is managed through a change of variables
mapping the star to the spherical geometry. Internal precision is
monitored through spectral convergence, the virial test and the
energy test \cite[see][for details]{REL12}. Iterations follow the
Newton algorithm.  Fig.~\ref{Massive} illustrates the result of a
two-dimensional model for a massive star rotating at 98\% of the critical
angular velocity. We show here the differential rotation as a function
of radius and latitude.

\begin{table}[t]
\caption{Comparison between observationally derived parameters of the
stars $\alpha$ Oph, $\alpha$ Lyr and $\alpha$ Leo and ESTER models. Data
are respectively from \cite{monnier_etal10}, \cite{monnier_etal12}
and \cite{che_etal11}. Note the good matching of the models.}
\vspace*{10pt}

\begin{tabular}{lllllll}

\hline
Star                     & \multicolumn{2}{c}{Ras Alhague ($\alpha$ Oph)} &
\multicolumn{2}{c}{Vega ($\alpha$ Lyr)} & \multicolumn{2}{c}{Regulus ($\alpha$
Leo)}\\
                         & Obs.     & Model  & Obs.     & Model &
Obs.     & Model\\
                         &                  &        &                  &       &
&       \\
Mass  (M$_\odot$)        & $2.4^{+0.23}_{-0.37}$ & 2.22 & 2.15$^{+0.10}_{-0.15}$
& 2.374 & 4.15$\pm$0.06  & 4.10  \\
R$_{\rm eq}$ (R$_\odot$) & 2.858$\pm$0.015  & 2.865& 2.726$\pm$0.006 & 2.726 &
4.21$\pm$0.07  & 4.24\\
R$_{\rm pol}$ (R$_\odot$)& 2.388$\pm$0.013  & 2.385& 2.418$\pm$0.012 & 2.418 &
3.22$\pm$0.05  & 3.23\\
T$_{\rm eq}$ (K)         & 7570$\pm$124     & 7674 & 8910$\pm$130    & 8973 &
11010$\pm$520  & 11175\\
T$_{\rm pol}$(K)         & 9384$\pm$154     & 9236 & 10070$\pm$90    & 10070 &
14520$\pm$690  & 14567\\
L (L$_\odot$)            & 31.3$\pm$1       & 31.1& 47.2$\pm$2      & 48.0 &
341$\pm$27     & 351\\
V$_{\rm eq}$ (km/s)      & 240$\pm$12       & 242  & 197$\pm$23      & 205 &
336$\pm$24     & 335\\
P$_{\rm eq}$ (days)      &                  & 0.598&                 & 0.672 &
& 0.641\\
P$_{\rm pol}$ (days)     &                  & 0.616&                 & 0.697&
& 0.658  \\
X$_\mathrm{env.}$&                          & 0.70 &     & 0.7546&       & 0.70\\
X$_\mathrm{core}$/X$_\mathrm{env.}$&        & 0.37 &     & 0.271 &     & 0.5\\
Z                &                          & 0.02 &     & 0.0093&     & 0.02\\
\hline
\end{tabular}
\label{rasalhague}
\end{table}

\subsection{Gravity darkening law}

One of the first results of steady models has been the prediction of the
gravity darkening law. Until now, the standard recipe was the von Zeipel
law stating that the effective temperature is proportional to the
$\frac{1}{4}$ power of the effective gravity. Fitting brightness
distributions, interferometric data of rapidly rotating stars have shown
that other laws like $T_{\rm eff}\propto g_{\rm eff}^{\beta}$ with
adjusted $\beta$, were more appropriate. This has also been the
conclusion of ESTER models with predictions on the $\beta$ value,
although the models show that a power law is not exactly representing
the gravity darkening \cite[][]{ELR11,ELR12}. In Fig.~\ref{flux}, we show  the
dependence of the flux with colatitude for a 30~\msun\ star rotating close to
critical angular velocity. The von Zeipel law underestimate the flux in the
equatorial regions by more than a factor 3.

\subsection{Some models of nearby fast rotating stars}

Fundamentals parameters of some nearby (famous) rotating stars,
derived by interferometry, have been compared successfully
to ESTER models as shown by Tab.~\ref{rasalhague}. A further
comparison is shown in Tab.~\ref{deltavel}. $\delta$ Vel A is an
eclipsing binary that has been studied in detail by interferometry
\cite[][]{merand_etal11,pribulla_etal11}. Here too, two dimensional
models nicely fit the observationnally derived parameters of the two
stars\footnote{The tidal distortion is weak, less than 10$^{-4}$.}. In
addition they suggest that they are less metallic than the Sun (Z$\simeq$
0.011). We also note the stronger hydrogen depletion in the core of the
most massive one as expected from its fastest evolution.

On the other hand the case of Achernar, the closest Be star to the solar
system, is more difficult since none of the explored models really fit the
parameters derived from interferometry. Fitting the polar and equatorial
radii lead to rather extreme composition or mass suggesting that the
star may just have left the main sequence (to which we are constrained
at the moment). In view of previous results based on spherical models
\cite[][]{vinicius_etal06}, this not totally surprising. However, this
star remains a challenging case for two-dimensional models.

\begin{table}
\caption{Comparison between observationally derived
parameters of the stars and tentative two-dimensional models. Data from
$\delta$ Vel are from \cite{merand_etal11}, those of Achernar are
from \cite{domiciano_etal12}. The models compare nicely with
observationally constrained data for the two components of $\delta$ Vel A
(an eclipsing binary) but have difficulties with Achernar.}
\vspace*{10pt}
\begin{tabular}{lllllll}
\hline
Star                     & \multicolumn{2}{c}{Delta Velorum Aa} &
\multicolumn{2}{c}{Delta Velorum Ab} & \multicolumn{2}{c}{Achernar ($\alpha$
Eri)}\\
                         & Obs.     & Model  & Obs.     & Model &
Obs.     & Model\\
                         &                  &        &                  &       &
&       \\
Mass  (M$_\odot$)        & $2.43\pm0.02$ & 2.43 & 2.27$\pm0.02$ & 2.27 &            & 8.20  \\
R$_{\rm eq}$ (R$_\odot$) & 2.97$\pm$0.02 & 2.95 & 2.52$\pm$0.03 & 2.52 &
11.6$\pm$0.3   & 11.5\\
R$_{\rm pol}$ (R$_\odot$)& 2.79$\pm$0.04 & 2.77 & 2.37$\pm$0.02 & 2.36 &
8.0$\pm$0.4  & 7.9 \\
T$_{\rm eq}$ (K)         & 9450          & 9440 & 9560          & 9477 &
9955$^{+1115}_{-2339}$  & 11250\\
T$_{\rm pol}$(K)         & 10100         & 10044 & 10120        & 10115 &
18013$^{+141}_{-171}$  & 16800\\
L (L$_\odot$)            & 67$\pm$3      & 65.2 & 51$\pm$2      & 48.5 &
4500$\pm$300   & 3700\\
V$_{\rm eq}$ (km/s)      & 143           & 143  & 150           & 153 &
298$\pm$9     & 339\\
P$_{\rm eq}$ (days)      &               & 1.045&               & 0.832 &
& 1.72 \\
P$_{\rm pol}$ (days)     &               & 1.084&               & 0.924 &
& 1.68   \\
X$_\mathrm{env.}$        &               & 0.70 &           & 0.70 &    & 0.74\\
X$_\mathrm{core}$/X$_\mathrm{env.}$&     & 0.10 &           & 0.30 &    & 0.05\\
Z                        &               & 0.011&           & 0.011&    & 0.04\\
\hline
\end{tabular}
\label{deltavel}
\end{table}

\section{Outlooks}

Presently, ESTER two dimensional models are, strictly speaking, models of
internal structure of rotating stars. No time evolution is included.  They
describe main sequence early-type stars, that is for masses larger than 1.8
M$_\odot$. Evolution along the main sequence can be mimicked
by varying the hydrogen mass fraction in the convective core. Clearly,
the next important steps are the extension to low mass stars, which
means the computation of outer convective envelope, and the inclusion
of time evolution. Other two-dimensional models are currently being
developed by Deupree and coworkers \cite[][]{deupree11,deupree_etal12},
but in these models the differential rotation should be prescribed.

Such internal structure models are useful to interpret the seismic
frequencies of rotating stars when perturbative methods fail. Presently,
two codes may deal with two dimensional models: TOP by Reese
\cite[][]{RLR06,reese_etal13} or ACOR by Ouazzani \cite[][]{ouazzani_etal12}.
Beside asterosismology, 2D-models are important for interferometry.
Indeed, the interpretation of interferometric visibilities requires the
adjustement of models that include the centrifugal distortion of the
stars as well as the associated gravity darkening \cite[e.g.][]{DVJJA02}.
Here, a future improvement will be the calculation of two-dimensional
models of atmospheres.

Beyond the obvious improvements that have been mentioned above, we see
that progresses in the understanding of rotating stars will have to go
through a better modeling of the turbulent transport in all places where
turbulence develops. This is certainly the most challenging issue of
this modeling since we do not have a general theory of turbulence at
hands.

\begin{acknowledgements}
I would like to warmly thank Francisco Espinosa Lara who has so much
contributed to the success of the ESTER project and with whom I had so
many enlighting discussions. The author acknowledges the support of
the French Agence Nationale de la Recherche (ANR), under grant ESTER
(ANR-09-BLAN-0140).  This work was also supported by the Centre National
de la Recherche Scientifique (CNRS, UMR 5277), through the Programme
National de Physique Stellaire (PNPS). The numerical calculations were
carried out on the CalMip machine of the ``Centre Interuniversitaire
de Calcul de Toulouse'' (CICT) which is gratefully acknowledged.
\end{acknowledgements}

%%-----------------------------
%%      your bibliography
%%-----------------------------
\bibliographystyle{aa}
%\bibliography{/home/virgo/tex/biblio/bibnew}
\bibliography{/home/rieutord/tex/biblio/bibnew}
\end{document}